\documentclass[
 reprint,
superscriptaddress,
 amsmath,amssymb,
 aps,
superscriptaddress,
 amsmath,amssymb,
 aps
]{revtex4-1}

\usepackage{graphicx}
\usepackage{dcolumn}
\usepackage{bm}
\usepackage{amsmath,amssymb,amstext,amsthm}
\usepackage{braket}
\usepackage{bbold}
\usepackage{bbm}
\usepackage{xcolor}
\usepackage{ gensymb }

\begin{document}

\preprint{APS/123-QED}
\setlength{\abovedisplayskip}{1pt}
\title{Generation of a lattice of spin-orbit beams via coherent averaging}

\author{D. Sarenac}
\email{dsarenac@uwaterloo.ca}
\affiliation{Department of Physics, University of Waterloo, Waterloo, ON, Canada, N2L3G1}
\affiliation{Institute for Quantum Computing, University of Waterloo,  Waterloo, ON, Canada, N2L3G1}
\author{D. G. Cory}
\affiliation{Institute for Quantum Computing, University of Waterloo,  Waterloo, ON, Canada, N2L3G1} 
\affiliation{Department of Chemistry, University of Waterloo, Waterloo, ON, Canada, N2L3G1}
\affiliation{Perimeter Institute for Theoretical Physics, Waterloo, ON, Canada, N2L2Y5}
\affiliation{Canadian Institute for Advanced Research, Toronto, Ontario, Canada, M5G 1Z8}
\author{J. Nsofini} 
\affiliation{Department of Physics, University of Waterloo, Waterloo, ON, Canada, N2L3G1}
\affiliation{Institute for Quantum Computing, University of Waterloo,  Waterloo, ON, Canada, N2L3G1}
\author{I. Hincks} 
\affiliation{Institute for Quantum Computing, University of Waterloo,  Waterloo, ON, Canada, N2L3G1}
\affiliation{Department of Applied Math, University of Waterloo, Waterloo, ON, Canada, N2L3G1}
\author{P. Miguel}
\affiliation{Institute for Quantum Computing, University of Waterloo,  Waterloo, ON, Canada, N2L3G1}
\affiliation{Department of Chemistry, University of Waterloo, Waterloo, ON, Canada, N2L3G1}
\author{M. Arif}
\affiliation{National Institute of Standards and Technology, Gaithersburg, Maryland 20899, USA}
\author{Charles W. Clark}
\affiliation{National Institute of Standards and Technology, Gaithersburg, Maryland 20899, USA}
\affiliation{Joint Quantum Institute, National Institute of Standards and Technology and University of Maryland, College Park, Maryland 20742, USA}
\author{M. G. Huber}
\affiliation{National Institute of Standards and Technology, Gaithersburg, Maryland 20899, USA}
\author{D. A. Pushin}
\email{dmitry.pushin@uwaterloo.ca}
\affiliation{Department of Physics, University of Waterloo, Waterloo, ON, Canada, N2L3G1}
\affiliation{Institute for Quantum Computing, University of Waterloo,  Waterloo, ON, Canada, N2L3G1}

\date{\today}


\pacs{Valid PACS appear here}


\begin{abstract}
We describe a highly robust method, applicable to both electromagnetic and matter-wave beams, that can produce a beam consisting of a lattice of orbital angular momentum (OAM) states coupled to a two-level system. We also define efficient protocols for controlling and manipulating the lattice characteristics. These protocols are applied in an experimental realization of a lattice of optical spin-orbit beams.  The novel passive devices we demonstrate here are also a natural alternative to existing methods for producing single-axis OAM and spin-orbit beams. Our techniques provide new tools for investigations of chiral and topological materials with light and particle beams.
\end{abstract}
\maketitle

Since their experimental demonstrations a quarter-century ago \cite{Bazhenov1990, LesAllen1992}, there has been great progress in generation, detection, and applications of ``structured waves'' of light and quantum particles, where the wavefront is patterned to attain nontrivial propagation characteristics such as orbital angular momentum (OAM), non-diffraction, and self-healing \cite{Harris2015,garces2002simultaneous,siviloglou2007observation,BarnettBabikerPadgett,molina2007twisted}. Beams of light \cite{ LesAllen1992}, 
neutrons \cite{Dima2015,holography} and electrons \cite{uchida2010generation,mcmorran2011electron} can carry orbital angular momentum parallel to their propagation axis. Furthermore, lattices of optical OAM beams have been produced and studied \cite{courtial2006angular,vyas2007interferometric,wei2009generation,masajada2007creation}. The structured OAM waves have demonstrated a number of applications in microscopy, encoding and multiplexing of communications, and manipulation of matter \cite{padgett2011tweezers,mair2001entanglement,wang2012terabit}.

Of particular interest are ``spin-orbit'' beams where the orbital degree of freedom is coupled to a two-level system such as polarization for light or spin for electrons and neutrons. These beams have found applications in high resolution optical imaging, high-bandwidth communication, and optical metrology \cite{rubinsztein2016roadmap,marrucci2011spin,milione20154}. Spin-orbit states of light beams may be prepared by an interferometric method using a spatial light
modulator \cite{maurer2007tailoring}, or via q-plates \cite{qplate}. The latter method is similar to preparing spin-orbit states via a space-variant Wien filter for electrons \cite{karimi2012spin} or via a quadrupole magnetic field for neutrons \cite{spinorbit}. 

The utility of the spin-orbit beams may be enhanced by producing a periodic lattice of such states, the lattice constants of which are matched to characteristic length scales of target materials. Here, we describe a universal parallel multiplexing technique that can produce a beam consisting of a lattice of OAM states coupled to a two-level system. Our protocols use coherent averaging and spatial control methods borrowed from nuclear magnetic resonance \cite{zhang1995analysis,cory1989chemical,sodickson1998generalized,levitt2007composite} to prepare a general pulse sequence for producing the lattices. Spin and polarization enter here as natural manifestations of the two degrees of freedom of light and spin$-1/2$ particles. The approach could be extended to systems with more degrees of internal freedom, such as atoms with higher spin. 

\begin{figure*}
\center
\includegraphics[width=\linewidth]{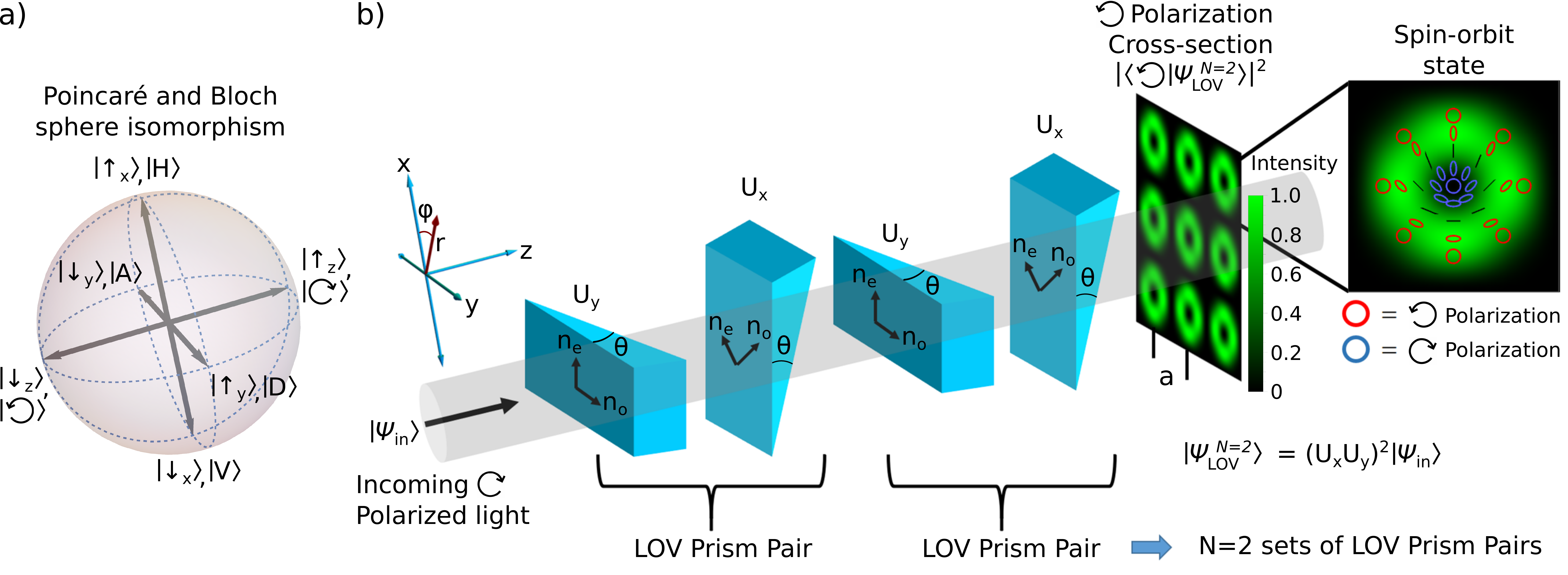}
\caption{a) The isomorphism between the Bloch sphere representing the spin states of fermions $\{\uparrow_x,\downarrow_x,\uparrow_y,\downarrow_y,\uparrow_z,\downarrow_z\}$ 
and that of the Poincar\'e\ sphere representing the polarization states of light $\{H,V,D,A,\circlearrowright,\circlearrowleft\}$. The corresponding eigenvectors are chosen as shown to ensure that $(r,\phi)$ are the transverse coordinates of the incoming beam. b) The lattices of optical spin-orbit beams are produced by passing a circularly polarized light beam through N sets of Lattice of Optical Vortices (LOV) prism pairs. A LOV prism pair consists of two perpendicular optical birefringent prisms where one prism has the optical axis along the prism incline and the second prism has the optical axis offset by $45\degree$. The lattice constant is given by $a=\lambda/2(n_e-n_o)\tan(\theta)$, where $\theta$ is the prism incline angle, $n_e$ and $n_o$ are the extraordinary and ordinary refractive indices, and $\lambda$ is the wavelength of the incoming light.}
 \label{Fig:Method}
\end{figure*}


To describe the protocols for creating and optimizing the lattices of spin-orbit beams, we first analyze a single spin-orbit state. It is convenient to consider a (light or particle) wavepacket traveling along the $z$-direction with momentum $\hbar k_z$ and expectation values of momentum in the transverse $(x,y)$ plane equal to zero. When the transverse coherence lengths are equal $\sigma_x=\sigma_y\equiv \sigma_\bot$, where $\sigma_{x,y}=1/(2\Delta k_{x,y})$, and $\Delta k_{x,y}$ are the $x$ and $y$ spreads of the wavepacket's transverse momentum distributions, the eigenstates in cylindrical coordinates $(r,\phi)$ can be expressed as

\begin{align}
	\ket{n_r,\ell,p}
    	= \mathcal{N} \xi^{|\ell|}e^{-\frac{\xi^2}{2}}
        	\mathcal{L}_{n_r}^{|\ell|}\left(\xi^2\right)e^{i\ell\phi}Z(z)\ket{p},
\label{basis}
\end{align}
where 
$\mathcal{N}=\frac{1}{\sigma_\perp}\sqrt{\frac{n_r !}{\pi(n_r+|\ell|)!}}$ is the normalization constant, $\xi=r/\sigma_\perp$ is the dimensionless radial coordinate, $\phi$ is the azimuthal coordinate, 
$n_r\in\{0,1,2...\}$ is the radial quantum number, $\ell\in \{0, \pm 1, \pm2...\} $ is the azimuthal quantum number, $\mathcal{L}_{n_r}^{|\ell|}\left(\xi^2\right)$ are the associated Laguerre polynomials, $Z(z)$ is the longitudinal wavefunction often approximated by a Gaussian wavepacket, and $p\in\{\circlearrowright,\circlearrowleft\}$ is the polarization state of light (or as per Fig.~\ref{Fig:Method}a we may use $s\in\{\uparrow_z,\downarrow_z\}$ in the case of spin$-1/2$ particles). Applying the OAM operator $\hat{L}_z=-i\hbar\frac{\partial}{\partial\phi}$ shows that the wavepacket carries an OAM of $\ell \hbar$. The coherence length $\sigma_\bot$ is important when dealing with particle beams where the beam is generally an incoherent mixture of coherent wavepackets, whereas for light one may simply consider the beam waist and the Laguerre-Gaussian modes. However, although the polarization-orbit beam can cleanly be described via Laguerre-Gaussian modes, the beam carrying a lattice of polarization-orbit states can not due to the translational symmetry.

When considering beams carrying OAM of major importance is the one fixed axis in space about which the OAM is quantized. In the case of beams carrying a lattice of OAM states there is a two-dimensional array of such axes and we are interested in what happens locally within each cell. Particularly, when this beam interacts with a material then the region around the local OAM axes becomes important.

To prepare states with coupled polarization and OAM we can start with circularly polarized light,

\begin{align}
\ket{\psi_{in}}=\ket{{0,0},\circlearrowright},
\label{Eqn:initialPsi}
\end{align}
and apply a coupling operator of the form \cite{spinorbit}

\normalsize
\begin{subequations} 
	\begin{align}
\hat{U}
    	&= e^{i\frac{\pi r}{2r_c}[\cos(\phi)\hat{\sigma}_x+\sin(\phi)\hat{\sigma}_y]} 
        \label{Eqn:UQ2} \\
 		&= \cos\left(\frac{\pi r}{2r_c}\right) \mathbb{1}
        	+i\sin\left(\frac{\pi r}{2r_c}\right)
            	\left(\hat{l}_+\hat{\sigma}_- +\hat{l}_-\hat{\sigma}_+\right).
	\end{align}
	
\end{subequations}
\normalsize
\noindent
Here, $\hat{l}_\pm =e^{\pm i\phi}$ are the raising and lowering OAM operators, $\hat{\sigma}_x$ and $\hat{\sigma}_y$ are the Pauli operators, and $\hat{\sigma}_\pm =(\hat{\sigma}_x \pm i\hat{\sigma}_y)/2$.
The length $r_c$ is defined as the smallest radial distance at which the polarization degree of freedom undergoes a $\pi$-rotation. At radii different than $r=r_c$, other rotation angles will occur producing the spin-orbit state

\small
\begin{align}
	\ket{\Psi_{SO}}
    	=\frac{e^{-\frac{r^2}{2}}}{\sqrt[]{\pi}}
        	\left[
            	\cos\left(\frac{\pi r}{2r_c}\right) \ket{\circlearrowright}
                +ie^{i\phi}\sin\left(\frac{\pi r}{2r_c}\right)\ket{\circlearrowleft}
            \right],
	\label{Eqn:psiq}
\end{align}
\normalsize
where we have set $\sigma_\bot=1$. $\ket{\Psi_{SO}}$ describes a vector vortex beam where the OAM is induced via Pancharatnam-Berry geometrical phase \cite{pancharatnam1956generalized,berry1987adiabatic}. The polarization distribution and the intensity post-selected on the right circularly polarized light is depicted on the zoomed-in plot of Fig.~\ref{Fig:Method}b. It has been shown that there is a correlation between the two degrees of freedom whereby post-selecting on one degree of freedom determines the value for the other, and that this correlation is maximized under the condition $r_c=1.82\sigma_\bot$ \cite{spinorbit}.

\begin{figure*}
\center
\includegraphics[width=.9\linewidth]{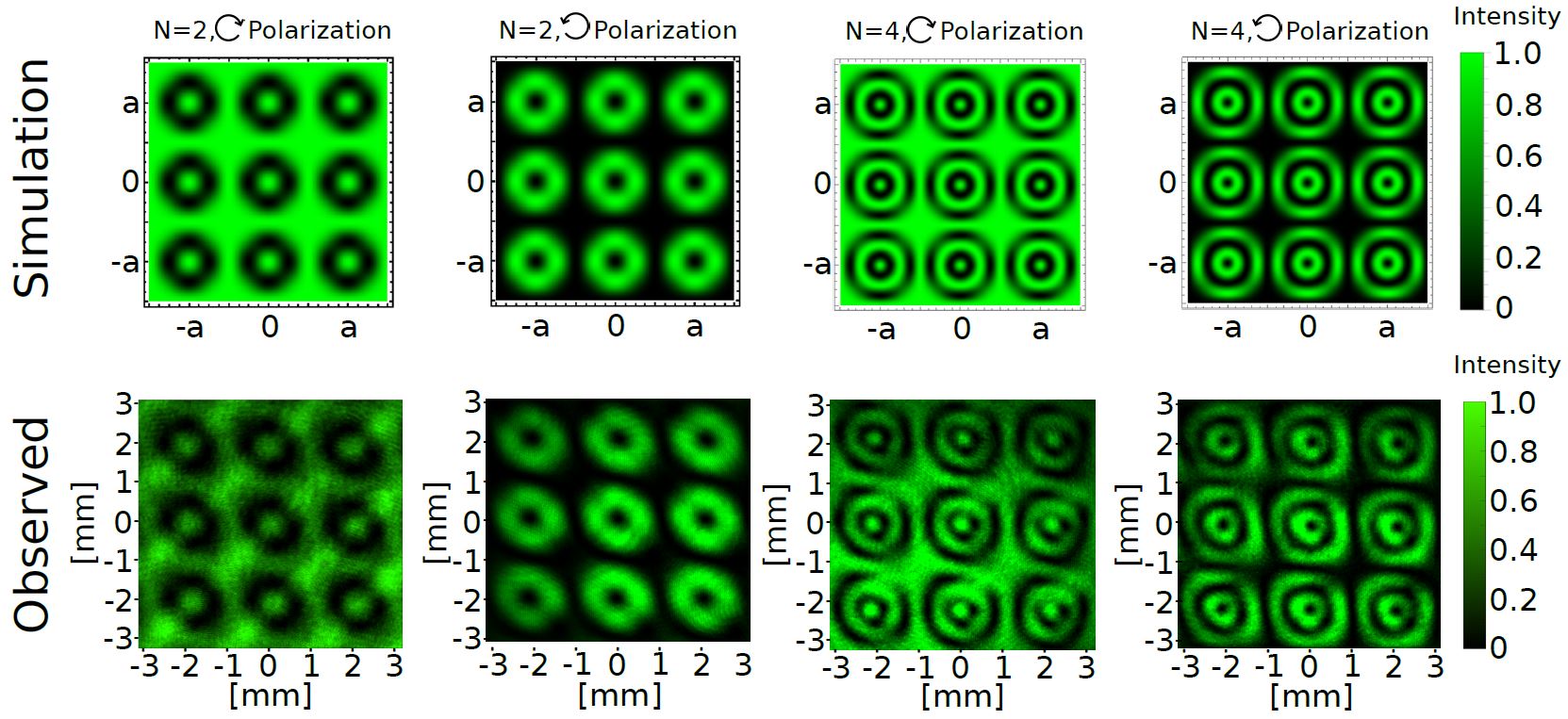}
\caption{Intensity profiles post-selected on a particular polarization state of the lattices of optical spin-orbit beams. Top are the simulated profiles and the bottom are the observed images. The lattice constant specified by Eq.~\ref{Eqn:const} for $\lambda=532$ nm light and our $2\degree$ quartz LOV prism sets is $a=1.68$ mm; the measured lattice constant at the camera being slightly larger due to beam divergence. If desired, the lattice constant can easily be pushed into the $\mu$m-range by fabricating prisms  with a larger incline angle out of a high birefringent material such as TiO$_2$.  }
 \label{Fig:PolarizationProfiles}
\end{figure*}

Our proposed procedure for producing a beam with a lattice of optical spin-orbit states consists of a sequence of linear birefringent gradients that are equal in magnitude and perpendicular to each other and the polarization axis of the incoming light. 
This procedure may be motivated by applying the Suzuki-Trotter expansion to Eq.~\ref{Eqn:UQ2}, i.e.

\begin{align}
	e^{i\frac{\pi}{2r_c}(x\hat{\sigma}_x+y\hat{\sigma}_y)}
    	= \lim_{N\to\infty} 
        	(e^{i\frac{\pi}{2r_c}x\hat{\sigma}_x/N}
            e^{i\frac{\pi}{2r_c}y\hat{\sigma}_y/N})^N,
	\label{Eqn:Lie}
\end{align}
where we have switched from radial to Cartesian coordinates,
$x=r \cos(\phi)$ and $y=r\sin(\phi)$.
Examining and truncating the right hand side of this relation, we see that it can be interpreted as a sequence of $N$ perpendicular linear gradients. 
Generalizing to put the origin of the gradients at $(x_0,y_0)$ and choosing that the gradients be independent of N, we define the operators

\begin{align}
	\hat{U}_x=e^{i\frac{\pi }{2 r_c}(x-x_0)\hat{\sigma}_x} 
    \qquad 
    \hat{U}_y=e^{i\frac{\pi }{2 r_c}	(y-y_0)\hat{\sigma}_y}
	\label{gradientoperators}
\end{align}
In the case of photons one way to produce the operators is via optical birefringent prisms as shown in Fig.~\ref{Fig:Method}b. 
Placing one prism with an optical axis along the prism incline and a second prism with an optical axis offset by $45\degree$ results in the product operation $\hat{U}_x \hat{U}_y$. We term such a set a ``Lattice of Optical Vortices (LOV) prism pair''. Eq.~\ref{gradientoperators} shows that a physical shift by a distance, $d$, of a prism along its incline direction (x or y) results in a simple phase shift  of $d\pi/2r_c$ around the corresponding axis. A sequence of $N$ sets of LOV prism pairs generates a lattice of optical spin-orbit beams, calculated as

\begin{align}
	\ket{\Psi_{LOV}^N}
    	=(\hat{U}_x \hat{U}_y)^N\ket{\psi_{in}}.
	\label{Eqn:psigrads}
\end{align}

\begin{figure*}
\center
\includegraphics[width=\linewidth]{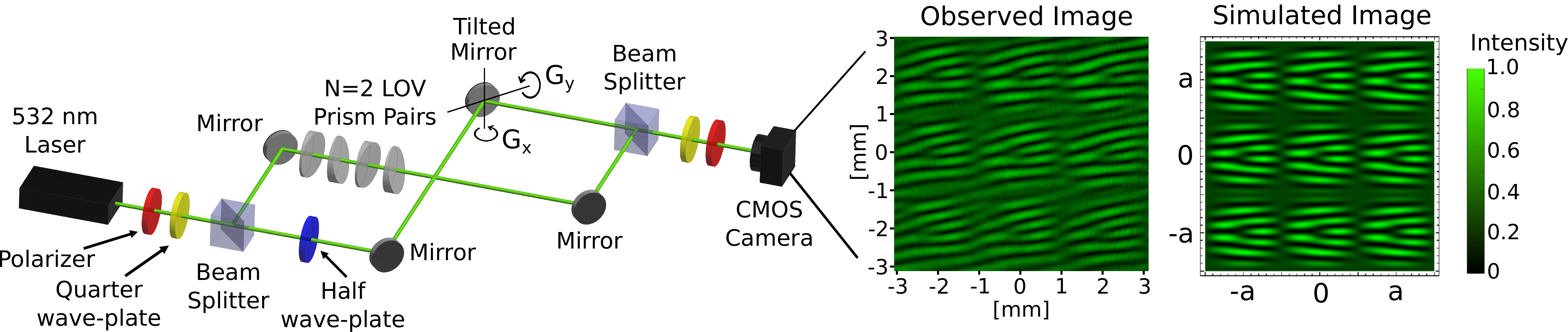}
 \caption{Phase imaging of the $N=2$ lattice of optical spin-orbit beams where we postselect on the polarization carrying the OAM. The $N=2$ sets of LOV prism pairs are placed in one path of the interferometer and a linear phase gradient is applied in the other path ($G_y\sim20$ rad/mm) by tilting a mirror in order to pronounce the well known fork structure holograms in the lattice, which indicate the presence of OAM beams.
 }
 \label{Fig:PhaseImaging}
\end{figure*}

This process is shown in Fig.~\ref{Fig:Method}b for $N=2$. The spin-orbit states in these lattices form a two-dimensional array with a lattice constant of

\begin{align}
	a=2 r_c=\frac{\lambda}{\Delta n \tan(\theta)}
	\label{Eqn:const}
\end{align}

\noindent where $\Delta n$ and  $\theta$ are the birefringence and the incline angle of the LOV prism pairs. The OAM structure of the resulting beam can be analyzed by looking at the phase profile of the polarization state which is correlated with the OAM:

\begin{align}
\arg\left(\braket{\circlearrowleft|\Psi_{LOV}^N}\right)=-\tan^{-1}\left[\cot\left(\frac{\pi y}{a}\right)\tan \left(\frac{\pi x}{a}\right)\right].
\label{Eqn:phase}
\end{align}
By analyzing Eq.~\ref{Eqn:phase} it can be observed that the lattice cells are centered on a $\ell_z=1$ phase structure, while the lattice cell corners are on a $\ell_z=-1$ structure. Although the number ($N$) of LOV prism pairs does not affect the phase profile, in any lattice cell the number of well defined intensity rings is equal to $N/2$. Therefore, N provides control over the mean radial quantum number $n_r$ in a lattice cell, and even expansions of Eq.~\ref{Eqn:Lie} should be used. 
In the $N=1$ case both polarization states are similarly coupled to the OAM, and both $\ell_z=1$ and $\ell_z=-1$ phase structures are illuminated. Similar vortex-antivortex structures can also be obtained via Wollaston prisms \cite{kurzynowski2006optical,kurzynowski2010regular}.

The simulated and observed polarization profiles for $N=2$ and $N=4$ are plotted in Fig.~\ref{Fig:PolarizationProfiles}, and are in a good agreement. For our LOV prism pairs the lattice constant  given by Eq.~\ref{Eqn:const} comes out to be $1.68$ mm, though it was measured to be slightly larger due to beam divergence.

The period of the lattice can span a large range. LOV prism pairs fabricated from TiO$_2$ (birefringence of $\sim0.29$) with an incline angle of $60\degree$ would produce a lattice period of $a\sim 1$ $\mu$m for a light wavelength of 532 nm. Furthermore, if birefringent materials which exhibit the Pockel's effect are used then with the addition of external electric field control a variable period may be obtained via the electro-optic effect. 

The doughnut structure in the spin-orbit states shown in  Fig.~\ref{Fig:PolarizationProfiles} is indicative of the polarization profile of the polarization-orbit state (Eq.~\ref{Eqn:psiq}) and not due to the OAM structure. To show that there is a lattice of OAM states we measure the phase profile of the beam using an interferometer. The schematic of the setup is shown in Fig.~\ref{Fig:PhaseImaging} where a linear phase gradient in one path has been introduced to observe the characteristic fork structure hologram indicative of OAM. A lattice of fork structures can clearly be seen, indicating an $\ell_z=1$ at each lattice center. 

Lattices of $\ell_z=-1$ spin-orbit states may be obtained by orienting the first prism of the LOV prism pairs along the negative $y$-direction. While various sequences of LOV prism pairs and polarization filters may be used to achieve higher order OAM structures in the outgoing beam. For example, to increment the OAM values to which the polarization states are coupled to by an integer ``m'', the following sequence may be used: 

\begin{align}\label{Eqn:higherq}
\left((\hat{U}_x \hat{U}_y)^Ne^{-i\frac{\pi}{2}\hat{\sigma}_x}\ket{\circlearrowleft}\bra{\circlearrowleft}\right)^{m-1}(\hat{U}_x \hat{U}_y)^N\ket{\psi_{in}}
\end{align}
\noindent where $\ket{\circlearrowleft}\bra{\circlearrowleft}$ is the operator for a polarization filter along the $\ket{\circlearrowleft}$ direction. Lastly, using LOV prism pairs which produce different lattice constants results in a ``superlattice'' which has an overlay of the distinct lattice constants.


The described protocols provide a two-dimensional control of the characteristic length scale of the single  spin-orbit features. It may be possible to create a lattice of ring-shaped optical atomic traps, individual instance of which have figured prominently in recent studies of atomtronic circuits \cite{eckel2014hysteresis}. One can also envisage vortex pinning in Bose-Einstein condensates via these beams \cite{tung2006observation}. Lattices of polarization coupled optical vortices may also be fruitful in microscopy or basic studies of the interaction of structured light \cite{Andersen2006,he1995direct,friese1996optical,brullot2016resolving} with individual atoms or molecules. This is because OAM is defined with respect to a single axis perpendicular to the wavefront. Thus, in studies using a single OAM axis, only atoms or molecules in the region of a fraction of a wavelength about that axis are subject to the OAM selection rules \cite{afanasev2016high,schmiegelow2016transfer}. This technique extends those rules across a region proportional to the area of the fully-structured wavefront.

Our technique is particularly useful for matter-wave beams where the beam is generally an incoherent mixture of coherent wavepackets. In the case of spin$-1/2$ particles, to create a lattice of spin-orbit states one requires a magnetic prism set with the magnetic field along the direction of the prism incline, and where the prisms are perpendicular to each other and the spin state of the incoming particles. Matter-wave lattices of spin-orbit beams may thus be generated where the OAM axis is specified along the coherent wavepacket rather than the beam axis. This opens the door for new types of studies of chiral and topological materials via particle beams.

\section{\label{sec:level1}Methods}
A laser of wavelength 532 nm was used, along with standard polarizers, wave-plates, and optical components. The LOV prism pairs were circular quartz wedges (birefringence of $\sim 0.0091$) with a wedge angle of $2\degree$ and diameter of $2.54$ cm. One wedge had the optical axis aligned with wedge angle while the other wedge had the optical axis aligned $45\degree$ to wedge angle.

For images shown in Fig.~\ref{Fig:PolarizationProfiles} the setup consisted of a laser, a linear polarization filter, a quarter-wave plate, $N$ LOV prism pairs, a quarter-wave plate, a linear polarization filter, and a CMOS camera. For beam phase imaging shown in Fig.~\ref{Fig:PhaseImaging}, a four-mirror interferometer was used because it allowed for compensation of the beam deviation due to the LOV prism pairs. An alternative method would have been to add a non-birefringent prism after each prism of the LOV prism pair in order to compensate for the beam deviation. A linear phase gradient in Fig.~\ref{Fig:PhaseImaging} was introduced to obtain the fork structure holograms by tilting the mirror of the interferometer path which did not contain the LOV prism pairs.

\section{Acknowledgements}

This work was supported by the Canadian Excellence Research Chairs (CERC) program, the Natural Sciences and Engineering Research Council of Canada (NSERC) Discovery program, Collaborative Research and Training Experience (CREATE) program, the Canada  First  Research  Excellence  Fund  (CFREF), and the National Institute of Standards and Technology (NIST) Quantum Information Program. The authors thank Connor Kapahi for his contributions to the figures.

\bibliography{OAM}

\end{document}